\documentclass[12pt]{iopart}

\usepackage{epsf}  
\usepackage{amssymb}  

\def\binomial#1#2{\left({#1 \atop #2}\right)}

\begin{document}

\title[Emergence of effective dynamical constraints]{Glassiness
through the emergence of effective dynamical constraints in
interacting systems}

\author{
Juan P. Garrahan\footnote{E-mail address: j.garrahan@physics.ox.ac.uk}
}

\address{Department of Physics, Theoretical Physics, University of
Oxford, 1 Keble Road, Oxford OX1 3NP, United Kingdom}

\begin{abstract}
I describe a class of spin models with short--range plaquette
interactions whose static equilibrium properties are trivial but
which display glassy dynamics at low temperatures. These models have a
dual description in terms of free defects subject to effective kinetic
constraints, and are thus an explicit realization of the constrained
dynamics picture of glassy systems.
\end{abstract}

\section{Introduction}

Perhaps the simplest systems which display the slow cooperative
relaxation characteristic of glasses are the facilitated kinetic Ising
models, first introduced in \cite{Fredrickson}, in which glassiness is
not a consequence of either disorder or frustration in the
interactions, but of the presence of kinetic constraints in the
dynamics of the system. Depending on the nature of the constraints
these models may behave as strong \cite{Fredrickson,Schulz,Crisanti}
or fragile \cite{Jackle,Sollich} glasses, display stretched
exponential relaxation \cite{Palmer,Buhot}, fragile-to-strong
transitions \cite{Buhot}, etc.

Kinetically constrained systems are interesting for several reasons.
First, they are usually free of disorder or frustration, and the
interactions imposed by the constraints are short-ranged. Second,
their equilibrium static properties are trivial, so that their
glassiness is a purely dynamical effect, in contrast to mean-field
models \cite{Kirkpatrick} where statics and dynamics are closely
related \cite{Franz}. Third, their low temperature behaviour is
dominated by activated processes, which, although highly relevant for
supercooled liquids near the glass transition, are not taken into
account by mode-coupling theory \cite{Gotze} or in mean-field
\cite{Bouchaud}. Finally, the simplicity of these models may allow for
exact solutions or at least a thorough analytical description of their
behaviour, which can help to understand dynamically arrested systems in
general.

The rationale behind the kinetically constrained approach was most
clearly given in \cite{Palmer}: it should be possible to describe the
dynamics of a strongly interacting system which displays glassy
behaviour in terms of hierarchies of degrees of freedom, from fast to
slow, independent of the presence of disorder or even frustration;
these hierarchies would be weakly interacting in the energetic sense,
but their dynamics should be constrained, the faster modes
constraining the slower ones. Systems with explicit kinetic
constraints, like the spin facilitated models or the constrained
lattice gases \cite{Kob,Kurchan}, represent one side of this picture,
but concrete examples in which this full scenario is realized are much
harder to find. The purpose of this paper to address this issue by
describing a simple class of models of interacting spins with `normal'
dynamics which have a dual description in terms of free excitations
whose dynamics is subject to effective kinetic constraints. They are
thus an explicit realization of the constrained dynamics scenario of
\cite{Palmer}.

\section{Models}

Consider the following model: a system of Ising spins $\sigma=\pm1$ on
a triangular lattice, where each spin interacts in triplets with its
nearest-neighbours lying at the vertices of a downward-pointing
triangles \cite{Newman,Garrahan}
\begin{equation}
H_3 = -J \sum_{ijk \in \nabla} \sigma_i \, \sigma_j \, \sigma_k 
\label{H3}
\end{equation} 
This system belongs to a class of models of Ising spins interacting
with its neighbours through plaquette interactions, such that the
spins and bonds live in isomorphic lattices and there is one bond per
spin. The simplest example is the one-dimensional Ising model (see
e.g. \cite{Baxter}). In two dimensions, in addition to model
(\ref{H3}), there is the plaquette model on the square lattice
\cite{Lipowski}
\begin{equation}
H_4 = -J \sum_{ijkl \in \square} \sigma_i \, \sigma_j 
	\, \sigma_k \, \sigma_l
\label{H4}
\end{equation} 
which is a special case of Baxter's eight vertex model \cite{Baxter}.
These models can be generalized to higher dimensions. Four spins
interacting in the corners of downward pointing tetrahedra of an fcc
lattice would be a generalization of (\ref{H3}), while eight spin
interactions in the vertices of cubes of a simple cubic lattice that
of (\ref{H4}). The sign of the couplings is irrelevant for large
systems, and even disordered $\pm J$ can be absorbed by a rescaling of
the spins. In what follows $J>0$ is assumed. 

Below I will mainly focus on the triangular model of Eq.\ (\ref{H3}),
but it will be instructive to compare with the 1d Ising model and the
square plaquette model (\ref{H4}).

\section{Spin-defect duality}

The common feature of the class of models of the previous section is
that they all have, for appropriate boundary conditions, a one-to-one
mapping between spin and bond (or `defect') configurations.  For the
Ising model on the line this happens if one boundary is fixed, say,
$\sigma_0=1$, and the other free. In this case the spin to bond
mapping and its inverse simply reads
\begin{equation}
b_i \equiv \sigma_{i-1} \, \sigma_i \, \Rightarrow \sigma_i =
\prod_{j=1}^i b_j 
\label{b1}
\end{equation}
Similarly, for the square plaquette model there is a $1$--$1$ mapping if
the spins on one boundary row and column are fixed, say,
$\sigma_{i0}=\sigma_{0i}=1$, and the other boundary
is left free
\begin{equation}  
b_{ij} \equiv \sigma_{i-1 j-1} \, \sigma_{i j-1}
	\, \sigma_{i-1 j} \, \sigma_{ij}
\Rightarrow \sigma_i = \prod_{k,l=1}^{i,j} b_{kl} 
\label{b4}
\end{equation}
Notice that, while the bonds are defined locally in terms of its
surrounding spins, the inverse mapping is nonlocal. In order to derive
it I made implicit use of the effect that introducing an excitation
has in the ground state configuration $\sigma=1$, i.e., how the spin
configuration has to be rearranged to avoid any further defects.

For the more interesting case of the triangular plaquette model the
situation is more complicated. The spin-to-bond mapping is again
straightforward
\begin{equation}
b_{ij} \equiv \sigma_{ij} \, \sigma_{i j+1} \, \sigma_{i-1 j+1}
\label{b3}
\end{equation}
where the indices $i$ and $j$ run along the unit vectors of the
triangular lattice $\vec{a}_1 \equiv \hat{x}$ and $\vec{a}_2 \equiv
\frac{1}{2} (\hat{x} + \sqrt{3} \hat{y})$. The inverse map can be
obtained from the following observation \cite{Martin}. A ground state
of the system can be constructed row by row: the state of one row of
spins determines the following one. A row configuration can be
specified by a dipolynomial (i.e., a generalization of a polynomial
with terms with both positive and negative powers of the argument)
$P(z)=\sum_n (1- 2 \sigma_n) \, z^n$, where $\sigma_n$ are the spins
on the row, and $n$ runs from $-L/2$ to $L/2$, where $L$ is the
horizontal length of the system. The following row $\sigma'$ is then
given by $P'(z) = \hat{T} P(z)$, where $\hat{T}=1+z$ is the operator
which propagates rows, and the coefficients in $P'$ are taken modulo
2. Suppose a defect is created at position $n=0$ in a ground state of
all spins up.  The row just below the defect has a spin down at $n=0$
and its dipolynomial is $P(z)=1$. Then the $m$--th row below it will
be given by $P^{(m)}(z)=\hat{T}^m =(1+z)^m \, {\rm mod} \, 2$, which
means that the propagation of the original spin down produces other
down spins at the positions for which the combinatorial numbers
$\binomial{n}{m}$ mod~2 are nonzero, a structure known as a Pascal
triangle (PT), see Fig.\ 1.  The spin configuration due to a
configuration of defects is obtained by the superposition mod~2 of the
PTs that each individual defect generates and the ground state on
which the excitation has been produced.

\begin{figure}
\begin{center}
\leavevmode
\epsfxsize=2.5in
\epsffile{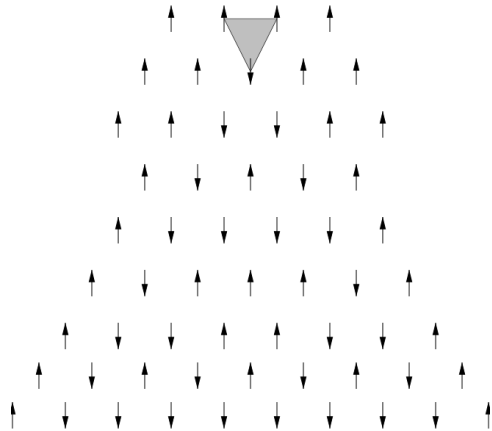}
\caption{The Pascal triangle of down spins formed by the presence of a
defect, shown as a shaded triangle, in the triangular plaquette model.}
\end{center}
\end{figure}

If the system is of linear length a power of two and has periodic
boundary conditions in at least one direction then there is a unique
ground state given by all spins up, and the spin-bond mapping is
1--1 \cite{Newman}. Moreover, the dipolynomial $P(z)$ has to be taken
modulo $1+z^L$ (spins $\sigma_1$ and $\sigma_{L+1}$ are identified),
and, since $\hat{T}^L \, {\rm mod} \, 2 = 1+z^L$, the propagation
described above stops at the $L$--th row. For this case Eq.\
(\ref{b3}) can be inverted: the spin at site $ij$ is given by the
superposition of the PTs of all the defects $b_{kl}=-1$ with $i-l\le
k\le i$ and $l\ge j$,
\begin{equation}
\sigma_{ij} =  
\prod_{l=j}^{j+L-1} \prod_{k=i-l}^{i} b_{kl}^{\binomial{l-j}{i-k}} 
\label{s3}
\end{equation}

Equation (\ref{s3}) is analogous to Eqs.\ (\ref{b1}) and (\ref{b4}). An
important difference is that, while for the 1d Ising and square
plaquette models a spin in the bulk is the product of $O(L^d)$
defects, for the triangular model it is given the product of
$O(L^{d_f})$, where $d_f=\ln{3}/\ln{2}$ is the fractal dimension of
the PT. Similar considerations apply to the higher dimensional models.

\section{Symmetries}

Consider now the symmetries of $H_3$. Suppose a whole row of spins is
inverted. None of the bonds below that row is affected since pairs of
spins on the row interact with spins in the next one. To leave the
bonds immediately above unchanged it is necessary to invert every other
spin in the preceding row. As one goes up the situation becomes more
complicated, but eventually it can be seen that if one inverts two PTs
of spins, as shown in Fig.\ 2, $H_3$ remains invariant.  The choice of
the initial row of flipped spins and the column from which the PTs
stem is arbitrary, as is the orientation of the transformation, which
can be in the three symmetry directions of the problem. This gives a
total of $3N$ transformations. However, a bit of algebra with binomial
coefficients shows that there are at most $2 L$ independent symmetry
operations. First, by combining transformations in two of the symmetry
directions, say $\vec{a}_1$ and $-\vec{a}_2$, gives a transformation
in the third direction, $\vec{a}_2 - \vec{a}_1$. Second, only the
choice of row is free, since the central column can be shifted by
superposition of transformations in different rows. 

The number of symmetries has to be the same as the number of ground
state configurations. By looking at the cycles of the operator
$\hat{T}$ \cite{Martin} it can be shown that in general there are at
most $O(L)$ ground states. For the special case of periodic boundary
conditions of length a power of two, for which the ground state is
unique, it is not difficult to prove that any transformation like that
of Fig.\ 2 `wraps' around on itself and becomes trivial. In that case
$H_3$ has no exact symmetries.

Once again, the symmetries of $H_3$ are a complicated analogue of the
ones in the 1d Ising model, which simply has the global inversion
symmetry, and the square plaquette model, where $H_4$ is left
invariant by the inversion of any row or column. In both these cases
they are exact symmetries only for periodic or free boundary
conditions.

\begin{figure}
\begin{center}
\leavevmode
\epsfxsize=3.5in
\epsffile{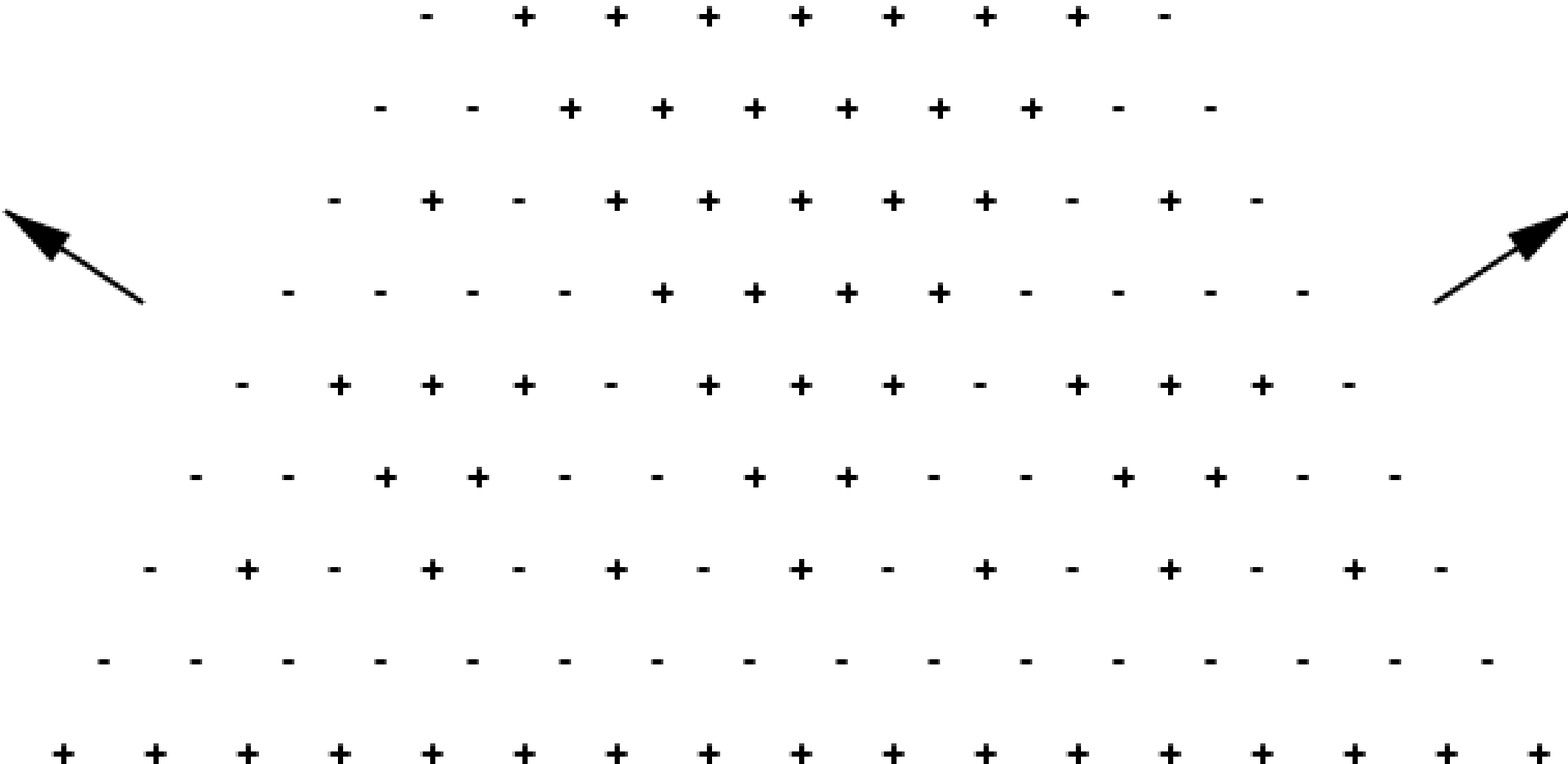}
\caption{Example of a symmetry transformation of $H_3$. Spins denoted
by `-' are inverted.}
\end{center}
\end{figure}

\section{Statics}

With the information of the two previous sections it becomes easy to
calculate the static properties of the models. If for each case the
boundary conditions are chosen such that the spin-defect mapping is
one to one, the bonds are noninteracting and the partition function is
simply that of a system of $N$ free Ising excitations, $Z_N = 2^N \,
\cosh^N(J/T)$. Now, for both the 1d Ising model and the square
plaquette model, we know from exact diagonalization of their transfer
matrices \cite{Baxter} that for arbitrary boundary conditions the
partition function per site in the thermodynamic limit is
\begin{equation} 
\kappa \equiv \lim_{N \to \infty} Z_N^{1/N} = 2 \, \cosh(J/T)
\label{kappa}
\end{equation} 
In the case of the triangular model, for linear size $L=2^K$ with
periodic boundary conditions, the row-to-row transfer matrix has a
single nonzero eigenvalue equal to $\kappa^L$. For arbitrary boundary
conditions it can be checked by numerical diagonalization that the
largest eigenvalue is still $\kappa^L$, so that in the thermodynamic
limit the partition function per site is given by (\ref{kappa})
irrespective of the boundaries. Probably this can be proved using the
exact methods of \cite{Baxter}.  Similar evidence suggests that for
the higher dimensional models Eq.\ (\ref{kappa}) holds too.

The expectation value of bonds is 
\begin{equation}
\langle b \rangle = \tanh(J/T)
\end{equation}
and the equilibrium energy density is, up to constants, just the
concentration $c$ of defects
\begin{equation}
c = \frac{1}{2} [1-\tanh(J/T)]
\end{equation}
Using Eqs.\ (\ref{b1})--(\ref{b3}) all spin correlation functions can
be calculated. Since in all models a spin in the bulk is the product
$O(N^\delta)$ bonds (the precise power depending on the model), in the
limit of $N \to \infty$, one obtains
\begin{equation} 
m \equiv \langle \sigma \rangle = 0 
\end{equation}
for all $T > 0$, i.e., all models are disordered at any finite
temperature.  

Arbitrary correlation functions are computed in a similar manner
\begin{equation} 
\langle \sigma_i \cdots \sigma_j \rangle = [\tanh(J/T)]
^{{\cal N}_{i \cdots j}}
\end{equation}
where ${\cal N}_{i \cdots j}$ is the number of bonds which enter in
the definition of the product of spins $\sigma_i \cdots \sigma_j$.
Only operators for which ${\cal N}_{i \cdots j}$ is finite in the
thermodynamic limit have non-vanishing expectation values. As usual
the 1d Ising model is the simplest, the product of two spins $\sigma_i
\, \sigma_j$ corresponds to the product of all the bonds in between,
giving the usual exponential decay of correlations with separation.
In the square plaquette model all two and three point spin
correlations vanish. The first nontrivial correlations are for
quartets of spins in the vertices of rectangles, which corresponds to
the product of all enclosed bonds.

In the case of the triangular plaquette model the simplest
non-vanishing correlations are for triplets of spins located at the
vertices of downward pointing equilateral triangles of side $2^k$,
which is given by the product of all the enclosed defects which also
belong to the corresponding PT,
\begin{equation}
C_{2^k}^{(3)} = \langle \sigma_{ij} \, \sigma_{ij+2^k} \,
                    \sigma_{i-2^k,j+2^k} \rangle 
        = [\tanh(J/T)]^{3^k}
\end{equation}
Notice the exact scaling relation $C_l^{(3)}=[C_1^{(3)}]^{l^{d_f}}$.

The fact that most of the correlation functions are zero is a
consequence of the multiple symmetries of these models and that there
is no symmetry breaking in the thermodynamics: only spin products
which are invariant under {\it all} the symmetry operations can have
nonzero correlations.

\section{Dynamics}

While the statics of the class of models being discussed here is
trivial their dynamics is not. Under a single spin-flip dynamics like
Glauber or Metropolis all of these models, except, of course, the 1d
Ising, become dynamically arrested due to the appearance of energy
barriers to the relaxation.

Consider the case of the triangular model. A single-spin flip
corresponds to flipping the three bond variables around it, see Fig.\
3a. This means that an isolated defect is locally stable since any
change in the spins around it creates a new pair of defects and thus
increases the energy. Moreover, only pairs of defects at a distance
$l=2^k$ can be relaxed by means of local spin moves, and the energy
barrier is $\Delta E=k$ (where we have set $J=1/2$), see Fig.\ 3b.  As
the system relaxes it has to cross barriers which grow as the
logarithm of the size of equilibrating regions.

\begin{figure}
\begin{center}
\leavevmode
\epsfxsize=3.5in
\epsffile{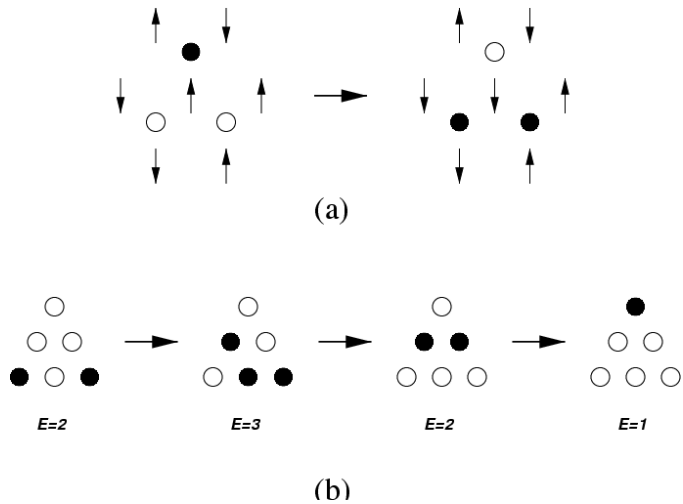}
\caption{(a) Example of a spin-flip move. Defects are represented as
dark circles. (b) Activated relaxation of a pair of defects at a
distance $l=2$.}
\end{center}
\end{figure}

From the observation that barriers grow logarithmically it is
straightforward to estimate the equilibration time.  The rate of
relaxation of an excitation of linear size $l$ is given by the
Arrhenius formula $\Gamma(l) \sim \exp(-\ln l/ T\ln2)$.  Thus, after
time $t$ the average linear distance between defects is $l \sim
t^{T\ln2}$. The equilibrium value of this distance at low $T$ is
$l_{\rm eq} = c^{-{1/2}} \sim \exp(1/2T)$, and hence the equilibration
time is
\begin{equation}
\tau \sim \exp\left(\frac{1}{2\,T^2\,\ln 2}\right)
\end{equation}
This supra-Arrhenius form is known as B\"assler equation
\cite{Bassler} and corresponds to fragile behaviour in the
classification of Angell \cite{Angell}. The absence of any finite
temperature singularity is consistent with the fact that model has no
finite-temperature phase transition.

The existence of a hierarchy of energy barriers implies that the
dynamics takes places in stages corresponding to the relaxation of a
typical lengthscale at each stage. This is clearly seen in the decay of
the concentration of defects after a quench from $T=\infty$ to a low
temperature, Fig.\ 4a.  After an initial $T$-independent exponential
relaxation corresponding to the (barrierless) removal of pairs of
neighbouring defects, the concentration displays ``plateaus'', which
are more pronounced the lower the temperature.  These plateaus are the
result of the system becoming trapped in locally stable
configurations, and correspond to the different stages of the
dynamics. 

The dynamical behaviour of the triangular plaquette model is almost
identical to that of the asymmetrically constrained Ising chain (or
East model) \cite{Jackle,Sollich}. In fact, the typical distance
between defects (and its the whole distribution) at each stage of the
dynamics can be computed approximately using the independent interval
method of Ref.\ \cite{Sollich}. Fig.\ 4b shows that the agreement
between theory and simulations is reasonably good
\cite{Garrahan}. This means that in terms of the defect variables, the
dynamics if effectively constrained as in the East model. The
existence of a dual description in terms of interacting spins with
standard dynamics and in terms of free defects subject to effective
kinetic constraints makes this model an explicit realization of the
constrained dynamics scenario \cite{Palmer}.

\begin{figure}
\begin{center}
\leavevmode
\epsfxsize=6.0in
\epsffile{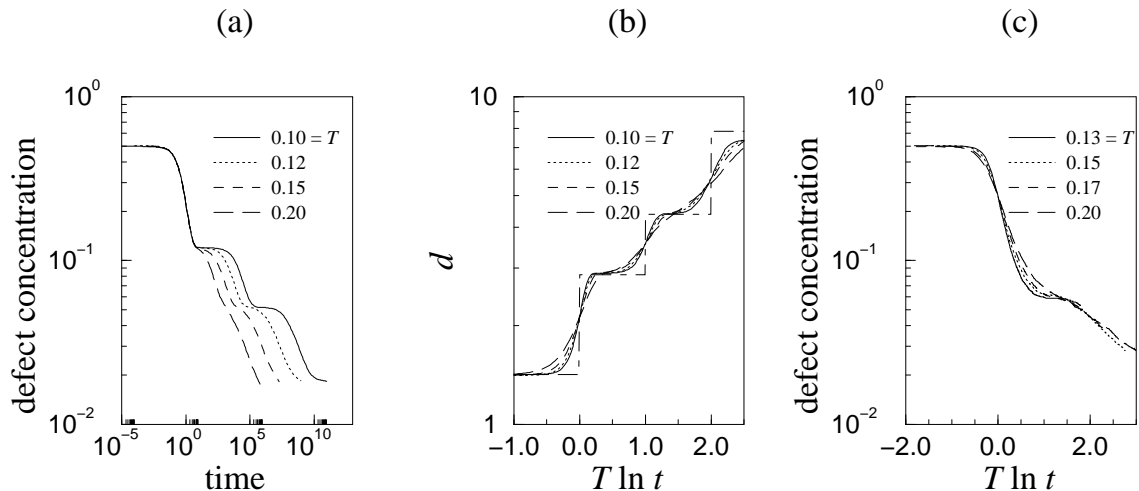}
\caption{(a) Concentration of defects as a function of time following
a quench from $T=\infty$ to various $T$, from MC simulations of model
(\ref{H3}). (b) Average distance between defects as a function of
rescaled time $T \ln t$.  The dot-dashed line corresponds to the
analytic approximation. (c) Decay of the concentration of defects as a
function of rescaled time in the 3--dimensional model of four spins
interacting in the corners of downward pointing tetrahedra of an fcc
lattice.}
\end{center}
\end{figure}

The dynamics of the square plaquette model is somewhat simpler,
although still activated. Since a pair of neighbouring defects can
diffuse freely, isolated excitations only face barriers of constant
size. Its glassy dynamics is strong, like the Fredrickson-Andersen
model \cite{Fredrickson,Crisanti}. The dynamics of the defects is
effectively given by the constrained lattice models of covalent
networks of Ref.\ \cite{Davison}. 

The generalization to three dimensions of the triangular model is that
of spins on an fcc lattice interacting with 4--spin interactions
between nearest--neighbours on the corners of downward pointing
tetrahedra. The energy barrier to the relaxation of excitations at a
distance $2^k$ is in this case $2\,k$.  In this model there are zero
energy moves when two defects are next to each other, but, in contrast
to the square model, these are not diffusive since every pair belongs
only to one plaquette. The relaxation time is of the B\"assler form,
$\tau \sim \exp\left(2/3\,T^2\,\ln 2\right)$, and the behaviour is
again fragile. Fig.\ 4c shows the decay of the concentration of
defects after a quench to low temperatures. Notice that the length of 
the plateaus is now of two units of rescaled time $T \ln t$. 

The symmetry considerations of the previous sections also apply to the
dynamics. Since the evolution operator has the same symmetries as the
Hamiltonian, dynamical expectation values of non-symmetric operators
will vanish if the initial configuration is taken from a distribution
which is also invariant, irrespective of whether the system is in
equilibrium or not, as in the out of equilibrium regime after a quench
from an initial infinite temperature state. This can be illustrated
with the triangular model. Figs.\ 5a and 5b show that all two--point spin
correlations functions vanish at all times after the quench, while the
invariant combination of triplets of spins in downward triangles of
length a power of two has a nontrivial behaviour in time. Fig.\ 5b
also shows how the stages of the dynamics correspond to the relaxation
of regions of size $2^k$.

\begin{figure}
\begin{center}
\leavevmode
\epsfxsize=2.5in
\epsffile{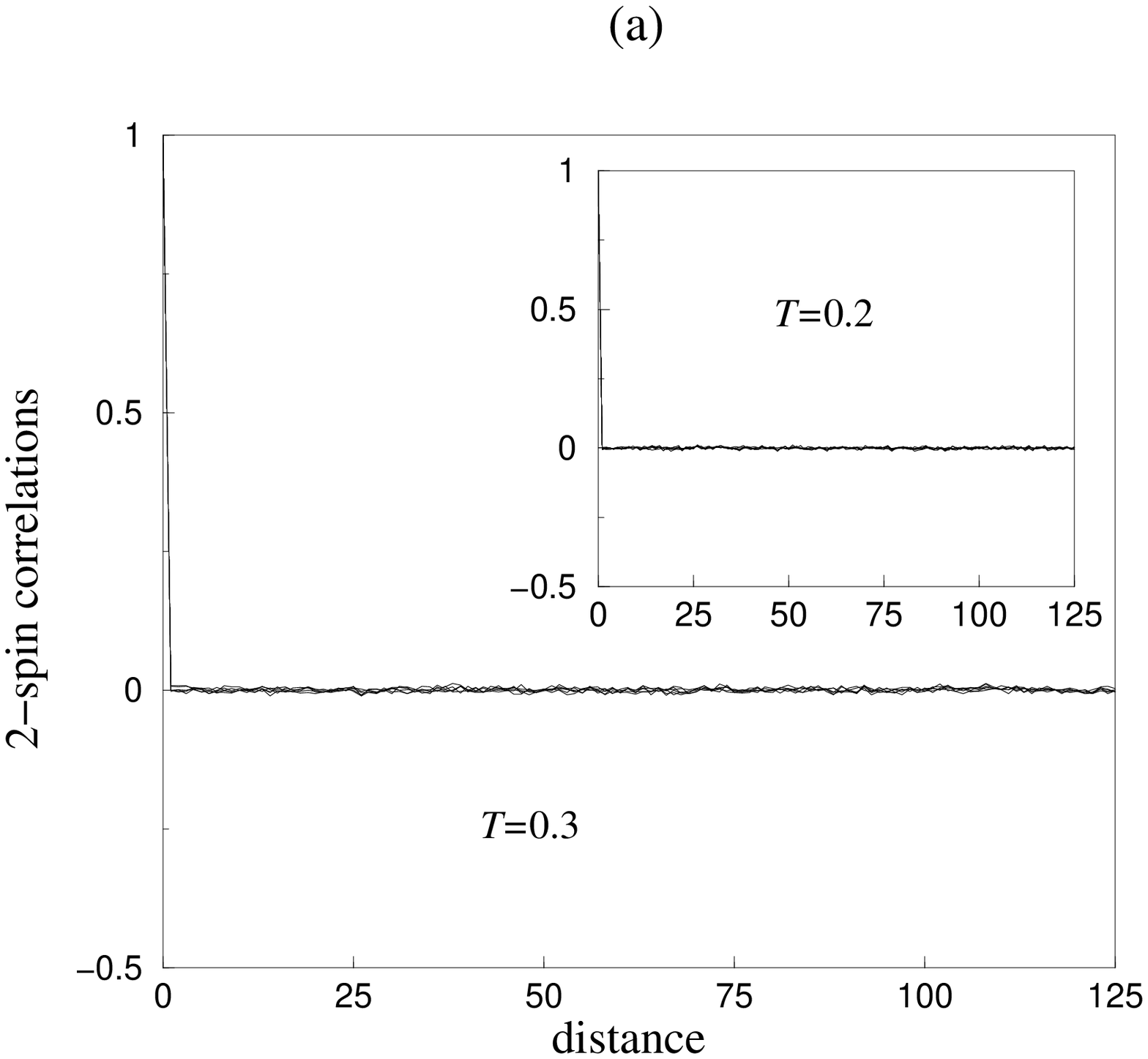}
\epsfxsize=2.5in
\epsffile{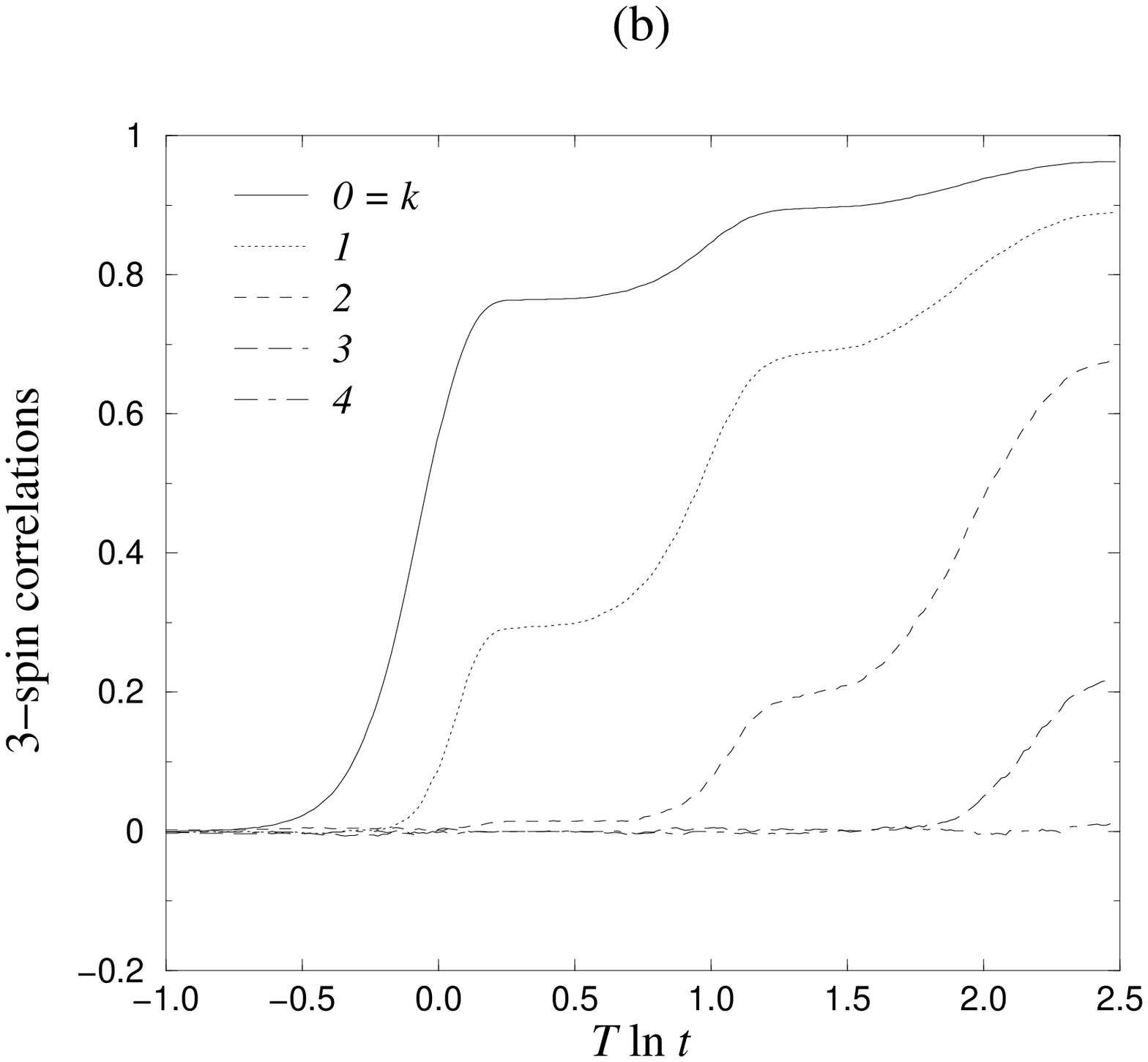}
\caption{(a) Two-spin correlations in the triangular model for times
$t=1$, $10$, $10^2$, $10^3$, $10^4$, and $10^5$ as a function of
distance. (b) Three-spin correlation functions $C_k^{(3)}$ as a
function of time, for linear sizes $2^{k=0,1,2,3,4}$, at $T=0.12$.}
\end{center}
\end{figure}

\section{Conclusions}

I have discussed a class of non--disordered and non--frustrated spin
models with local plaquette interactions which display glassy
behaviour due to the presence of energy barriers to the
relaxation. The low temperature dynamics is effectively that of free
excitations subject to dynamical constraints.

While rather unrealistic, these models have several interesting
features. They are explicit realizations of both sides of the
constrained dynamics scenario. Since the dynamics is defined through
their Hamiltonian, glassiness is not a consequence of {\it ad hoc}
dynamical rules, as in models with imposed kinetic constraints.  The
fact that their statics is trivial but their dynamics glassy is
consistent with recent observations in more realistic models
\cite{Santen}. And all the interesting behaviour happens in the
activated regime.

\ack

I warmly acknowledge the collaboration of Mark Newman in much of the
work described here.  I would also like to thank David Chandler and
Arnaud Buhot for important discussions. This work was supported by the
Glasstone Fund (Oxford).

\end{document}